\documentstyle[floats,pra,aps]{revtex}
\tightenlines

\def\beq{\begin{equation}}
\def\eeq{\end{equation}}
\def\bea{\begin{eqnarray}}
\def\eea{\end{eqnarray}}
\def\nn{\nonumber}
\def\ba{\begin{array}}
\def\ea{\end{array}}   
\def\i{{\rm i}}
\begin{document} 

\title{\Large\bf Quantum-like Approach to the Halo Formation in
High Current Beams}

\author{Sameen Ahmed KHAN}
\address
{E-mail: khan@pd.infn.it, ~~~ http://www.pd.infn.it/$\sim$khan/}

\author{Modesto PUSTERLA} 
\address
{
Dipartimento di Fisica Galileo Galilei  
Universit\`{a} di Padova \\
Istituto Nazionale di Fisica Nucleare~(INFN) Sezione di Padova \\
Via Marzolo 8 Padova 35131 ITALY \\
E-mail: pusterla@pd.infn.it, ~~~ http://www.pd.infn.it/$\sim$pusterla/}

\maketitle

\medskip
\noindent{\bf Abstract} \\
An interpretation of the formation of halo in accelerators based on
quantum-like theory by a diffraction model is given in terms of the
transversal beam motion. Physical implications of the longitudinal dynamics 
are also examined.



\section{Introduction}
A possible way of describing the evolution of a high density beam is
the so-called quantum-like 
approach~\cite{Fedele}, where one considers a time-dependent 
Schr\"{o}dinger equation, in both the usual linear and the less usual 
nonlinear form, as a fluid equation for the whole beam. In this case 
the squared modulus of the wave function~(named beam wave function) gives
the distribution function of the particles in space at a certain 
time~\cite{EPS}. The Schr\"{o}dinger equation may be taken in one or more
spatial dimensions according to the particular physical problem; 
furthermore the  motion of the particles in the configuration space can 
be considered as a Madelung fluid if one chooses the equation in its
linear version.

Although the validity of the model relies only on experiments and on the
new predictions, one may also invoke
a theoretical argument that could justify the Schr\"{o}dinger 
quantum-like approach. We notice that the 
normalized emittance~$\epsilon$ with the dimension of an action is the 
natural choice in the quantum-like theory, the analogue of
the Planck's constant~$\hbar$ because it reproduces the corresponding 
area in the phase-space of the particle.
After linearizing the Schr\"{o}dinger-like
equation, for beams in an accelerator, we can use the whole apparatus
of quantum mechanics, keeping in mind a new interpretation of the basic
parameters.
In particular one introduces the propagator 
$K \left( x_f , t_f | x_i , t_i \right)$ of the Feynman theory for both 
longitudinal and transversal motion. A procedure of this sort seems 
effective for a global description of several phenomena such 
as intrabeam scattering, space-charge, particle focusing, that cannot be 
treated easily in detail by ``classical mechanics''. One consequence of
this procedure is to obtain information on the creation of the {\em Halo}
around the main beam line by the losses of particles due to the transversal
collective motion. Here we shall mainly consider the
case of the HIDIF facility. The LHC has been discussed in~\cite{EPS}.

\section{Transversal Motion}
Let us consider the Schr\"{o}dinger like equation for the beam
wave function
\bea
\i \epsilon \partial _t \psi 
= - \frac{\epsilon^2}{2 m} \partial_x ^2 \psi + U \left( x , t \right) \psi
\label{schroedinger-like}
\eea
in the linearized case $U \left (x , t \right)$ does not depend on the 
density $\left| \psi \right|^2$. Here $\epsilon$ is the normalized
transversal beam emittance defined as follows 
$\epsilon = m_0 c \gamma \beta \tilde{\epsilon}$, where
$\tilde{\epsilon}$ is the emittance usually considered,
Let us now focus on the one 
dimensional transversal motion along the $x$-axis of the beam particles 
belonging to a single bunch and assume a Gaussian transversal profile 
for particles injected into a circular machine. We want to try a 
description of interactions that cannot be treated in detail, as a
diffraction through a slit that becomes a phenomenological boundary
in each segment of the particle trajectory.
The result is a multiple integral that determines the 
actual propagator between the initial and final states in terms of the 
space-time intervals due to the intermediate segments.
\bea
K \left(x + x_0 , T + \tau | x' , 0 \right) 
& = &
\int_{- b}^{+ b}
K \left(x + x_0 , \tau | x_0 + y_n , T + (n - 1) \tau ' \right) \nn \\
& & \quad \times 
K \left(x + y_n , T + (n - 1) \tau ' | 
x_0 + y_{n - 1} , T + (n - 2) \tau ' \right) \nn \\
& & \qquad \qquad \qquad \times \cdots 
K \left(x + y_1 , T | x' , 0 \right) d y_1 d y_2 \cdots d y_n 
\label{integral}
\eea
where~$\tau = n \tau '$ is the total time spent by the beam in the
accelerator~(total time of revolutions in circular machines), $T$ is the
time necessary to insert the bunch (practically the time between two
successive bunches) and $(-b , +b)$ the space interval defining the
boundary mentioned above. Obviously $b$ and $T$ are phenomenological 
parameters which vary from a machine to another and must also have a 
strict correction with the geometry of the vacuum tube where the particles
circulate.
We may notice that the convolution property~(\ref{integral}) of the
Feynman propagator allows us to substitute the multiple integral
(that becomes a functional integral for $n \longrightarrow \infty$ and 
$\tau ' \longrightarrow 0$) with the single integral
\bea
K \left( x + x_0 , T + \tau | x' , 0 \right) 
= \int_{- b}^{+ b} dy
K \left( x + x_0 , T + \tau | x_0 + y , T \right) 
K \left( x_0 + y , T | x' , 0 \right) dy
\label{single}
\eea
where $x_0$ is the initial central point 
of the beam at injection and can be chosen as the origin of
the transverse motion of the reference trajectory in the frame of the
particle. {\bf $\hbar$ must be interpreted as the 
normalized beam emittance in the quantum-like approach}.
With an initial Gaussian profile the beam wave function 
(normalized to 1) is
\bea
f (x) = \left\{ \frac{\alpha}{\pi} \right\}^{\frac{1}{4}}
\exp{\left[- \frac{\alpha}{2} x'^2 \right]}
\eea
$\sqrt{\frac{1}{\alpha}}$ being the r.m.s transversal spot size of the 
beam; the final beam wave function is:
\bea
\phi (x) 
= 
\int_{- \infty}^{+ \infty} d x'
\left(\frac{\alpha}{\pi} \right)^{\frac{1}{4}}
e^{\left[- \frac{\alpha}{2} x'^2\right]}
K \left(x, T + \tau ; x', 0\right) 
\label{bw}
\eea
We may consider the two simplest possible approximations for 
$K \left( n | n - 1 \right) \equiv 
K \left( x_0 + y_n , T + (n - 1) \tau ' | 
x_0 + y_{n - 1} + (n - 2) \tau ' \right)$:

\begin{enumerate}

\item
We substitute the correct $K$ with the free particle $K_0$ assuming that 
in the $\tau '$ interval $(\tau ' \ll \tau)$ the motion is practically a
free motion between the boundaries $( -b , + b )$.

\item
We substitute it with the harmonic oscillator 
$K_{\omega} \left( n | n -1 \right)$ considering the betatron and the 
synchrotron oscillations with frequency $\omega/{2 \pi}$

\end{enumerate}

The final local distribution of the beam that undergoes the diffraction is
therefore 
$\rho (x) = \left| \phi (x) \right|^2 
= B B^{*} \exp{\left[ - \tilde{\alpha} x^2 \right]}$
and the total probability per particle is given by
\bea
P = \int_{- \infty} ^{+ \infty} d x \rho ( x ) 
\label{probability}
\eea

\section{Longitudinal Motion}
As far as the longitudinal motion is concerned the quantum-like approach
appears to be quite appropriate to obtain information on the modified
length~(and consequently the stability) of the bunches both in the linear
and circular accelerators. 
We introduce the Gaussian parameter $b$, as we did with the Gaussian 
slit $e^{{-x^2}/{2 b^2}}$ in the transversal motion and look for a 
phenomenological solution of the equation for the beam wave 
function~$\psi$
\bea
\i \epsilon_N \partial_t \psi 
= - \frac{\epsilon_N^2}{2 \gamma^3 m_0} \partial_x^2 \psi 
+ \frac{1}{2} m_0 \omega^2 x^2 \psi + \Lambda \left| \psi \right|^2 
\label{schroedinger-nonlinear}
\eea
where $\omega$ is the synchrotron frequency, $\Lambda$ represents the 
coupling with non-linear terms and $x$ is the longitudinal particle
displacement with respect to the synchrotronous one.
Numerical calculations for the HIDIF project are shown in
Table~I. 

\section{Comments and Conclusions}

\noindent{\bf Transversal Motion}:
This use of a quantum-like approach appears a simple powerful tool for 
the analysis of the evolution of a beam in linear \& circular
accelerators and storage rings.
Indeed the introduction of a very limited number of phenomenological
parameters~(in our simplified model the only parameter $b$) in the
beam quantum-like equations and the use of the Schr\"{o}dinger-type
solutions allow us to calculate how the bunches evolve and modify
owing to the forces~(linear and non-linear) acting on the particles.

As far as the betatron oscillations are concerned the mechanism of the 
diffraction through a slit appears a very adequate phenomenological 
approach. Indeed we can interpret the probability~(local and total) for 
a particle leaving its position as the mechanism of creating a {\bf halo} 
around the main flux.


The phenomenological parameter $b$ represents several fundamental
processes that are present in the beam bunches~(and play a determinant
role in the creation of the halo) such as intrabeam scattering,
beamstrahlung, space-charge and imperfections in the  magnets of 
the lattice that could cause non-linear perturbative effects.

We like to recall here the analogy with the diffraction through 
a slit in optics where it represents a much more complicated physical 
phenomenon based on the scattering of light against atomic electrons.

We remark now the following points 

\begin{enumerate}

\item
The total probability~(per particle) calculated from the free particle
propagator~($P$) and from the harmonic oscillator one~($P_{\omega}$)
appear very near for the storage rings in HIDIF~(and LHC).

\item
The HIDIF scenario, as we expect because of the higher intensity,
exhibits a total loss of particles~(and beam power) which is at 
least $10^3$ times higher than LHC. The picture we have obtained for 
the transversal motion
is encouraging because the halo losses are under 
control. The estimated losses of the beam power
appear much smaller than the permissible $1$~Watt/m.

\end{enumerate}

\noindent
{\bf Longitudinal motion}
The formula~(\ref{bw}) can be used for calculating the motion 
of the length of the bunch related to the synchrotron oscillations in both 
linear and circular machines. In this case we must consider only the 
propagator of the harmonic oscillator which is the simplest linear version 
of the classical dynamical motion for the two canonical conjugate variables 
that express the deviations of an arbitrary particle from the synchronous 
one namely the RF phase difference 
$\Delta \phi = \phi - \phi_s$ and the energy difference 
$\Delta E = E - E_s$.
Our examples is the main LINAC in the HIDIF project. 
The phenomenological Gaussian function
$e^{- {x^2}/{2 b^2}}$ acquires a different meaning from the one it had in 
the transversal motion. Our analysis deals with a Gaussian longitudinal 
profile and predicts a coasting beam in LHC and a quite stable bunch in the
main LINAC of HIDIF.

We may therefore conclude that our approach although preliminary is
interesting and particular attention is required in treating the
longitudinal motion where the nonlinear space-charge forces are very
important. 

Numerical calculations for the HIDIF project are shown in
Table~I.

\begin{center}

{\bf TABLE-I: Circular Machines: Transversal Case}


\begin{tabular}{lll}
{\bf Parameters} & {\bf HIDIF} (main LINAC) & {\bf HIDIF} (storage ring) \\
Normalized Transverse Emittance ~~~~~~~
& $0.7$ KeV nano sec. ~~~~~~~~~~~ & $13.5$ mm mrad \\
Total Energy, $E$ & $5$ GeV & $5$ Gev \\
$\frac{1}{\sqrt{\alpha}}$ & $15$ cm & $1.0$ mm \\
$T$ & $75$ micro sec. & $100$ nano sec. \\
$\tau$ & $4.9 \times 10^{-4}$ sec. & $4.66$ sec. \\
$b$ & $15$ m & $1.0$ mm \\
$\frac{1}{\sqrt{\tilde{\alpha}}}$ & --- & 
$1..96 \times 10^{7}$ m \\
$P$ & --- & $2.37 \times 10^{-3}$ \\
$\omega$ & $4.13 \times 10^{5}$ Hz & $1.15 \times 10^{7}$ Hz \\
$\frac{1}{\sqrt{{\tilde{\alpha}_\omega}}}$ & $6.72 \times 10^{-2}$ m
& $2.07 \times 10^{-1}$ m \\
$P_{\omega}$ & $0.707$ & $3.00 \times 10^{-3}$ \\
\end{tabular}

\end{center}

\end{document}